\documentclass{ws-ijmpa}

\newcommand{\nn}{\nonumber}
\newcommand{\bd}{\begin{document}}
\newcommand{\ed}{\end{document}}
\newcommand{\bc}{\begin{center}}
\newcommand{\ec}{\end{center}}
\newcommand{\be}{\begin{eqnarray}}
\newcommand{\ee}{\end{eqnarray}}
\newcommand{\ba}{\begin{array}}
\newcommand{\ea}{\ed{array}}
\renewcommand{\thefootnote}{\alph{footnote}}
\newcommand{\se}{\section}
\newcommand{\sse}{\subsection}

\begin{document}

\markboth{Chen, Geng and Lih}
{Decay Spectrum of $K^+ \to e^+ \nu_{e}\gamma$}

%
\catchline{}{}{}{}{}
%

\title{DECAY SPECTRUM OF $K^+ \to e^+ \nu_{e}\gamma$  }

\author{ C.H. CHEN$^{1}$, C.Q. GENG$^{2}$
  and C.C. LIH$^{3}$}

\address{ $^{1}$Department of Physics, National Cheng-Kung
University, Tainan 701, Taiwan \\
$^{2}$Department of Physics, National Tsing-Hua University,
Hsinchu 300, Taiwan  \\
$^3$General Education Center, Tzu-Chi  College of Technology,
Hualien 970, Taiwan
}

\maketitle


\begin{abstract}
The  form factors of the $K^+ \to \gamma$ transition are studied
 in the light-front quark model and chiral perturbation theory 
of  $O(p^6)$.
The decay spectrum
of $K^+ \to e^+ \nu_{e}\gamma$,  dominated by the structure dependent  contribution,
is illustrated in both models.
\\
\\
\end{abstract}


It is known that the decay of $K^+\to e^+\nu_e\gamma$
 receives two types of contributions:  ``inner bremsstrahlung'' (IB) and ``structure-dependent'' (SD)
\cite{GW,Bryman}. The former is  helicity suppressed and
contains the electromagnetic coupling constant $\alpha$, while the latter 
 gives the dominant contribution to the decay rate as
 it is free of the helicity suppression. 
  In the standard model (SM), the decay amplitude
 of the SD part  involves
 vector and axial-vector hadronic currents, which can be parametrized in terms of the vector form factor $F_V$ and
 axial-vector form factor $F_A$, respectively.
 However, the experimental determinations on these form factors are
  poorly given and model-dependent \cite{75,79,E787}. In particular, the experimental results on the 
  decay rate of $K^+\to e^+\nu_e\gamma$ in Ref. \cite{75,79,E787} were based on the assumption of $F_V$ and $F_A$ being some constant values in
   the chiral perturbation theory (ChPT)
at $O(p^4)$ \cite{Bijnens93}.
   In the ongoing data analysis of the E949 experiment at BNL,
 more precision measurements on the decay of $K^{+}\to e^{+}\nu_{e}\gamma$ are expected \cite{E949}
 and thus, the model-independent extractions of the SD form factors are possible.
 Theoretical calculations of $F_V$ and $F_A$ in the $K^+\to \gamma$ transition
have been previously done
 in the 
ChPT
at $O(p^4)$ \cite{Bijnens93}
and $O(p^6)$ \cite{mod3,mod4}
as ell as the light-front quark model (LFQM) \cite{modLF}.
However, the results  have not been fully applied to
the decay of $K^{+}\to e^{+}\nu_{e}\gamma$ yet. 

In this talk, we will  present our recent results \cite{CGL} on
 the transition form factors of $K^+\to\gamma$ in the ChPT of $O(p^6)$ and light-front quark model (LFQM).
We will show the spectrum of 
the differential decay branching ratio of
$K^{+}\to e^{+}\nu_{e}\gamma$
as a function of $x=2E_\gamma/m_K$.

We start with
 the amplitude of the decay $K^{+}\to e^{+}\nu_{e}\gamma$
 in the SM, given by~\cite{Bryman,Bijnens93,ffdef}
\be
M&=&M_{IB}+M_{SD}, \nonumber
\\
M_{IB}&=&i e{G_F \over \sqrt{2}} sin\theta_c F_K m_{e}
 \epsilon^*_{\alpha}K^{\alpha},\ \;
M_{SD}=-i e{G_F \over \sqrt{2}} sin\theta_c \epsilon^*_{\mu}L_{\nu}H^{\mu\nu},
\label{eqn:sd}
\ee
where
$K^{\alpha}=\bar{u}(p_{\nu})(1+\gamma_5)\left ({p_K^{\alpha} \over p_K\cdot q}
-{2 p^{\alpha}_{e}+ \not\!q \gamma^{\alpha} \over 2 p_{e}\cdot q}
\right)v(p_{e})$,
$L_{\nu}=\bar{u}(p_{\nu})\gamma_{\nu}(1-\gamma_5)v(p_{e})$,
$ H^{\mu\nu}={F_A \over m_K}\left(-g^{\mu\nu}p_K\cdot q+p_K^{\mu}q^{\nu}\right)+
i{F_V \over m_K}\epsilon^{\mu\nu\alpha\beta}q_{\alpha}p_{K\beta}$
$\epsilon_{\alpha}$ is the photon polarization vector,
$p_K$, $p_{\nu}$, $p_{e}$, and $q$ are the four-momenta of
$K^+$, $\nu_e$, $e^+$, and $\gamma$,
and $F_K$ and $F_{A(V)}$
are the $K$ meson decay constant
and  the axial-vector
(vector) form factor corresponding to the axial-vector (vector) part of the weak currents,
respectively,
 defined by
\be
\langle\, 0|\bar{s}\gamma^{\mu}\gamma_5 u|K^+(p_K) \,\rangle &=&-iF_Kp_K^{\mu},
\
\langle\gamma(q) |\bar{u}\gamma^{\mu }s|K(p_K) \,\rangle =
ie{\frac{F_{V}}{m_{K}}}\varepsilon^{\mu \alpha \beta \nu }
\epsilon_{*\alpha }q_{\beta }p_\nu \, ,
\nn\\
\langle\gamma(q) |\bar{u}\gamma^{\mu }\gamma _{5}s|K(p_K) \,\rangle &=&
e{\frac{F_{A}}{%
m_{K}}}\left[ (p\cdot q) \epsilon ^{* \mu}
-(\epsilon ^{*}\cdot p)q^{\mu }\right] ,
\label{4}
\ee
with $p=p_K -q$ being the transfer momentum.
We note that ${\cal M}_{IB}$  in Eq. (\ref{eqn:sd}) is suppressed due to the small
electron mass $m_{e}$.
In the decay of $K^{+}\to e^{+}\nu_{e}\gamma$,
the form factors $F_{A,V}$ in Eq.  (\ref{4}) are the analytic functions of $p^{2}=(p_K -q)^2$ in
 the physical allowed region of
$m_{e}^{2}\leq p^{2}\leq m_{K}^{2}$. The relation between the transfer momentum $p^2$ and $x$ is given by 
$p^2=m_{K}^{2}(1-x)$.

 At $O(p^6)$ in the ChPT,
one obtains 
that \cite{CGL}
\be
F_{V}(p^2)=\frac{m_{K}}{4\sqrt{2}\pi^{2}F_K}\bigg\{1-\frac{256}{3}\pi^{2}m_{K}^{2}C_{7}^{r}
+256\pi^{2}(m_{K}^{2}-m_{\pi}^{2})C_{11}^{r}+\frac{64}{3}\pi^{2} p^{2} C_{22}^{r}
\nonumber\\
-\frac{1}{16 \pi^{2}(\sqrt{2}F_{K})^{2}}\bigg[\frac{3}{2}m_{\eta}^{2}
\ln \left(\frac{m_{\eta}^{2}}{\mu^{2}}\right)+\frac{7}{2} m_{\pi}^{2}
\ln \left(\frac{m_{\pi}^{2}}{\mu^{2}}\right)
+3m_{K}^{2} \ln \left(\frac{m_{K}^{2}}{\mu^{2}}\right)
\nonumber\\
-2\int \left[x m_{\pi}^{2}+(1-x) m_{K}^{2}-x(1-x)p^{2}\right]
\ln\left(\frac{x m_{\pi}^{2}+(1-x) m_{K}^{2}-x(1-x)p^{2}}{\mu^{2}}\right)dx
\nonumber\\
-2\int \left[x m_{\eta}^{2}+(1-x) m_{K}^{2}-x(1-x)p^{2}\right]
\ln\left(\frac{x m_{\eta}^{2}+(1-x) m_{K}^{2}-x(1-x)p^{2}}{\mu^{2}}\right)dx
\nonumber\\
-4\int m_{\pi}^{2}\,
\ln\left(\frac{m_{\pi}^{2}}{\mu^{2}}\right)dx
\bigg]\bigg\}\,,~~~~~~~~~~~~~~~~~~~~~~~~~~~~~~~~~~~~~~~~~~~~~~~~~~~~~~
\label{fvk}
\\
F_{A}(p^2)=\frac{4 \sqrt{2}m_{K}}{F_{K}}(L _{9}^{r}+L _{10}^{r})+
\frac{m_{K}}{6F_{K}^{3}(2\pi)^{8}}[142.65 (m_{K}^{2}-p^{2})-198.3]
~~~~~~~~~~~~
\nonumber \\
-\frac{m_{K}}{4\sqrt{2}F_{K}^{3}\pi^{2}}\left\{ (4L
_{3}^{r}+7L _{9}^{r}+7L _{10}^{r})m_{\pi }^{2}\ln \left(\frac{m_{\pi }^{2}}{m_{\rho }^{2}%
}\right)+3\,(L _{9}^{r}+L _{10}^{r})m_{\eta }^{2}\ln \left(\frac{m_{\eta }^{2}}{%
m_{\rho }^{2}}\right)\right.
  \nonumber \\
\left. +2\,(8L
_{1}^{r}-4L _{2}^{r}+4L _{3}^{r}+7L _{9}^{r}
+7L _{10}^{r})m_{K}^{2}\ln \left(\frac{m_{K}^{2}}{m_{\rho }^{2}}%
\right)\right\} ~~~~~
\nonumber \\
-\frac{4\sqrt{2}m_{K}}{3F_{K}^{3}}\left\{ 2m_{\pi
}^{2}(18y_{18}^{r}-2y_{81}^{r}-6y_{82}^{r}
+2y_{83}^{r}+3y_{84}^{r}-y_{85}^{r}+6y_{103}^{r})\right.
~~~ \nonumber
\\ +2m_{K}^{2}(18y_{17}^{r}+36y_{18}^{r}-4y_{81}^{r}
-12y_{82}^{r}+4y_{83}^{r}+6y_{84}^{r}+4y_{85}^{r}-3y_{100}^{r}
~~~
\nonumber \\ +\left.
6y_{102}^{r}+12y_{103}^{r}-6y_{104}^{r}+3y_{109}^{r})
+\frac{3}{2}(m_{K}^{2}-p^{2})(2y_{100}^{r}-4y_{109}^{r}+y_{110}^{r})\right\},
~~
\label{fak}
\ee
where 
$C^{r}_{i}$, $L^{r}_{i}$ and $y^{r}_{i}$
are the renormalized coupling constants.
  Note that the first terms in Eqs. (\ref{fvk}) and (\ref{fak}) correspond to $F_V$ and $F_A$ at $O(p^4)$
\cite{Bijnens93},  respectively.

In the framework of the LFQM \cite{modLF}, 
  we obtain \cite{CGL}
\be
F_{A}(p^2) &=&4m_{K}
        \int \frac{dz\,d^{2}k_{\bot }}{2(2\pi)^{3}}\Phi
        \left( z',k_{\bot }^{2}\right) {1\over 1-z'}
 \left\{\frac{2}{3}\frac{m_{u}-Ak_{\bot }^{2}\Theta }
        {m_{u}^{2}+k_{\bot }^{2}}+ \frac{1}{3}\frac{m_{s}+Bk_{\bot }^{2}
        \Theta}{m_{s}^{2}+
        k_{\bot}^{2}}  \right\}\,,   \nonumber \\
F_{V}(p^2) &=&8m_{K}
        \int \frac{dz\,d^{2}k_{\bot }}{2\left( 2\pi \right) ^{3}}\Phi
        \left( z',k_{\bot }^{2}\right) {1\over 1-z'}
                \nonumber \\
&&\left\{ \frac{2}{3}\frac{m_{u}-
        z'\left( m_{s}-m_{u}\right) k_{\bot }^{2}\Theta }{m_{u}^{2}
        +k_{\bot }^{2}}-\frac{1}{3}\frac{m_{s}+(1-z')(m_{s}-m_{u}) k_{\bot }^{2}
        \Theta }{m_{s}^{2}+k_{\bot }^{2}} \right\}\,,
\label{FFLFQM}
\ee
where the parameters and variables are defined in Ref. \cite{CGL}.

\begin{figure}[pb]
\centerline{\psfig{file=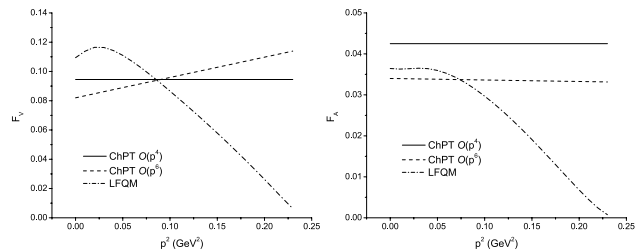,width=9.cm}}
\vspace{-3.3cm}
\caption{$F_{V,A}(p^2)$ as 
functions of the transfer momentum $p^2$. 
 \label{F2}}
\end{figure}

The numerical values of $F_{A,V}(p^2)$ in the ChPT of $O(p^6)$
are plotted  in Fig. \ref{F2}.
 In these figures, we have also included the results in the ChPT at $O(p^4)$.
Explicitly, we find that  $F_{V(A)}(p^2=0)=0.0945\ (0.0425)$, $0.082\ (0.034)$ and 
$0.106\ (0.036)$ in the ChPT at $O(p^4)$, ChPT at $O(p^4)$ and LFQM, respectively.

\begin{figure}[pb]
\centerline{\psfig{file=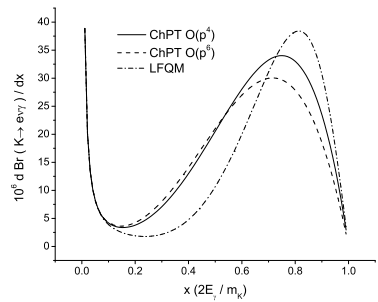,width=5.5cm}}
\vspace*{3.8pt}
\caption{The
differential decay branching ratio
as a function of $x=2E_{\gamma}/m_{K}$.
 \label{F5}}
\end{figure}

  The differential decay rate as a function of $x$ is given by
\be
\frac{d\Gamma }{dx}
&=&
\frac{m_{K}^{5}}{64\pi
^{2}} \alpha G_{F}^{2}\sin^{2}\theta_c  A(x)
\label{Rate1}
\ee
where the function of $A(x)$ is given in Ref. \cite{CGL}.
By integrating out the variable $x$  in Eq. (\ref{Rate1}), in Table \ref{Table2} we give the
decay branching ratio of $K^+ \to e^+ \nu_e \gamma$.
Here, as
the IB term diverges at the limit of
$x\to 0$ corresponding to $p^2\to p^2_{max}=m_K^2$, we have used the cuts of $x=0.01$ and $0.1$, respectively.
With the cuts, from Table \ref{Table2} we see that
the IB  contributions are much smaller than
 the SD$^\pm$ ones, which are insensitive to the cut.
  In Fig. \ref{F5}, 
we also display the spectrum of the differential decay
branching ratio  in the ChPT at both $O(p^4)$ and $O(p^6)$ and the LFQM.
\begin{table}[htbp]
 \tbl{The decay branching ratio  of $K^+ \to e^+ \nu_e \gamma$ (in units of $10^{-5}$).}
{\begin{tabular}{|c|c||c|c|c|c|c||} \hline
Model &$x$ Cut & IB & SD$^+$ & SD$^-$ & Total
\\ \hline \hline
ChPT at $O(p^4)$ & $0.01$&$ 1.65\times 10^{-1} $
& $ 1.34 $ & $1.93\times 10^{-1} $ &
 $ 1.70 $
\\
 &$0.1$& $ 0.69\times 10^{-1} $
& $ 1.34 $ & $1.93\times 10^{-1} $ &
$ 1.60 $
\\ \hline
ChPT at $O(p^6)$ &$0.01$& $ 1.65\times 10^{-1} $
& $ 1.15 $ & $2.58\times 10^{-1} $ & 
 $ 1.57 $
\\
& $0.1$&$ 0.69\times 10^{-1} $
& $ 1.15 $ & $2.58\times 10^{-1} $ &
 $ 1.47 $
\\ \hline
 LFQM & $0.01$&$ 1.65\times 10^{-1} $
& $ 1.12 $ & $2.59\times 10^{-1} $ &
$ 1.54 $
\\
 & $0.1$&$ 0.69\times 10^{-1} $
& $ 1.12 $ & $2.59\times 10^{-1} $ & 
 $ 1.44 $
\\ \hline
\end{tabular}
\label{Table2}}
\end{table}
 From Fig. \ref{F5},
we see that in the region of $x<0.7$ or $E_\gamma<173$ MeV,
 the decay branching ratio in the LFQM is much samller
than that in the ChPT at $O(p^6)$. On the other hand, in the region of $x>0.7$
the statement is reversed. However, if we only consider the contributions in the ChPT at $O(p^4)$,
the conclusion is weaker. 
It is clear in the future data analysis such as the one at the experiment BNL-E949 \cite{E949},
one could concentrate on these two regions to find out which model is preferred.


We have studied the axial-vector and vector form factors  of the $K ^+\to \gamma$ transition
in the LFQM and   ChPT of $O(p^6)$.
Based on these form factors, we have calculated the decay branching ratio of
 $K^+ \to e^+ \nu_{e}\gamma$. We have demonstrated that the SD parts give the dominant contributions
  to the decay in the whole allowed region of the photon energy except the low endpoint.
   Future precision experimental measurements on the decay spectrum \cite{E949}
should give us some useful information to
  determine the SD contributions as well as the vector and axial-vector form factors.




\begin{thebibliography}{0}    

\bibitem{GW}
J.T.~Goldman and W.J.~Wilson,
  Phys.\ Rev.\   {\bf D15}, 709 (1977).

\bibitem{Bryman}
D.A. Bryman {\em et al.}, Phys. Rep. {\bf 88}, 151 (1982).

\bibitem{75}
K.S.~Heard {\it et al.},
  Phys.\ Lett.\  {\bf B55}, 324 (1975).

\bibitem{79}
 J.~Heintze {\it et al.},
  Nucl.\ Phys.\   {\bf B149}, 365 (1979).

\bibitem{E787}
S. Adler {\em et al}. [E787 Collaboration], Phys. Rev. Lett. {\bf 85}, 2256 (2000).

\bibitem{Bijnens93}
J. Bijnens, G.Ecker, and J. Gasser, Nucl. Phys. {\bf B 396}, 81 (1993).

\bibitem{E949}
  D. A. Bryman, private communications.

\bibitem{mod3}
L. Ametller, J. Bijnens, A. Braman and F. Cornet, Phys. Lett. {\bf B303}, 140 (1993).

\bibitem{mod4}
C.Q. Geng, I.L. Ho and T.H. Wu, Nucl. Phys. {\bf B684}, 281 (2004).

\bibitem{modLF}
C.Q.~Geng, C.C.~Lih and W.M.~Zhang,
  Phys.\ Rev.\  {\bf D57}, 5697 (1998);
  $ibid$. {\bf D59}, 114002 (1999);
 $ibid$. {\bf D62}, 074017 (2000);
  Mod.\ Phys.\ Lett.\   {\bf A15}, 2087 (2000);
C.Q. Geng, C.C. Lih and C.C. Liu, Phys. Rev. {\bf D62}, 034019 (2000).

\bibitem{CGL}
  C.H.~Chen, C.Q.~Geng and C.C.~Lih, 
  Phys. Rev. {\bf D77}, 014004 (2008),
  arXiv:0710.2971 [hep-ph].

\bibitem{ffdef}
C.Q. Geng and S.K. Lee, Phys. Rev. {\bf D51}, 99 (1995);
 C.H.~Chen, C.Q.~Geng and C.C.~Lih,
  $ibid$. {\bf D56}, 6856 (1997).


\end{thebibliography}
\end{document}